\documentclass[aps,twocolumn,showpacs,showkeys,nofootinbib]{revtex4-1}
\usepackage{epsfig}
\usepackage{amsmath}
\usepackage{amsfonts}
\usepackage{amssymb}
\usepackage{graphicx}
\usepackage{colordvi}
\usepackage{color}
\usepackage{dcolumn}
\usepackage{multirow}
\usepackage{hyperref}
\usepackage{epstopdf}
\usepackage{bm}
\usepackage{times}
\hypersetup{
  colorlinks=true,        % false: boxed links; true: colored links
  linkcolor=blue,         % color of internal links
  citecolor=magenta,      % color of links to bibliography
}

\begin{document}

\title{Fluctuations of gravitational waves in Eddington inspired Born-Infeld theory}

\author{Celia Escamilla-Rivera}
\email{cescamilla@mctp.mx}
\affiliation {Mesoamerican Centre for Theoretical Physics,\\ Universidad Aut\'onoma de Chiapas.
Ciudad Universitaria, Carretera Zapata Km. 4, Real del Bosque (Ter\'an), 29040, Tuxtla Guti\'errez, Chiapas, M\'exico.}

%-------------------------------------------------------------------------------------------------------------------------------------------------
%-------------------------------------------------------------------------------------------------------------------------------------------------

\begin{abstract}
In this paper we review the EiBI gravity in the
presence of a cosmological constant and its tensor perturbations analysis. We show 
the existence of gravitational waves in the past-time, seeing as a result the smooth transition 
between high-energy densities (where the EBI dynamics plays its role) and low-energy densities (GR).
%Eddington and General Relativity regime. 
We obtain the fluctuation spectrum for the
graviton in this theory, where for small values of $k$ the fluctuations are strongly suppressed and for
large values of $k$ these fluctuations vanish during the De Sitter expansion.
\end{abstract}

%\keywords{Cosmology, Gravitational Waves, Graviton.}

%-------------------------------------------------------------------------------------------------------------------------------------------------
%-------------------------------------------------------------------------------------------------------------------------------------------------
        
\maketitle

\section{Introduction}

One of the greatest jigsaws in the current physics research is to
understand the nature of dark energy and dark matter \cite{Zhao:2017cud,Copeland:2006wr,
Bird:2016dcv,Clowe:2006eq}. 
%A detailed explanation of each one by its own is
%essential and very important for the entirely view of the universe. 
%The simplest way to model the dark sector in  good agreement with 
%observations of large scale structure of the universe is through the 
%$\Lambda$CDM model, which open the possibility to the existence of two 
%components: an energy dominant component with negative equation of state 
%and a ratio relative of $\Omega_{\Lambda}\approx 73\%$ and a matter 
%component $\Omega_m \approx 23\%$. 
Currently, dark energy is one of the main classes of models to describe the
cosmic late-time acceleration, which has been confirmed by a large number
of observations such as measurements SNIa \cite{Riess:1998cb-Perlmutter:1998np}, BAO \cite{Eisenstein:2005su},  CMBR anisotropies \cite{Spergel:2003cb}, LSS  \cite{Tegmark:2003ud} and WL \cite{Jain:2003tba}. Future projects and surveys \cite{surveys} are underway to discover the underlying cause of this phenomena.
Recently, the first multimessenger gravitational-wave (GW) observation of a binary neutron star made by LIGO-Virgo detector network set a way to infer cosmological parameters independently of the cosmic distance ladder \cite{TheLIGOScientific:2017qsa}, getting a better value for the Hubble constant -- and by extension, a better understanding of dark energy -- could be right on the horizon.

In the light of rich observed data, either we just know some properties of each component of the dark sector or one might have a new proposal of the gravitational theory without the need of these dark components instead.
Some attempts has been done in order to achieve these issues, e.g in \cite{Banados:2008fi} was presented a class of bigravity with solutions that can be interpolate between matter and acceleration epochs.
%new model that
%describes a coherent unified dark sector 
%where an Eddington Born-Infeld action (EBI) represent a connection
%between the fundamental fields. 
%so-called the affine connections, and
%the rest of the universe through the gravitation. 
%
%A remarkable
%point of this theory is the presence of two metrics, one of them $g_{\mu\nu}$
%is the usual metric that couples with matter. The second metric, $q_{\mu\nu}$,
%satisfies the Einstein-Hilbert action and couples to the rest of the 
%world through its interaction with $g_{\mu\nu}$. We view this theory as a bimetric model in where $q_{\mu\nu}$ can be
%interpreted as dark matter and not as an extra scalar field. The resulting cosmology
%show that the EBI scenario can mimic the standard cosmological
%evolution, where the field can behave as pressureless matter and cosmological 
%constant at the same time. These theories of bigravity are extensively 
%study in the literature \cite{Damour:2002ws,ArkaniHamed:2002sp,Blas:2005yk,
%Banados:2008fj} and reference therein.
%All these models precise an affine connection
%equivalent to Einstein gravity. To avoid this preference 
In \cite{Banados:2008fj,Banados:2010ix} was presented a non-conventional
formulation in terms of the affine connection $\Gamma^{\mu}_{\alpha\beta}$ and
a space-time metric $g_{\alpha\beta}$ such that the gravitational action is given by:
\begin{eqnarray}\label{palatini}
S_{EiBI}[g,\Gamma,\Psi]&=& \frac{2}{\kappa}\int{d^4 x \left[\sqrt{\left|g_{\mu\nu} 
+\kappa R_{\mu\nu}(\Gamma)\right|} -\lambda\sqrt{g}\right]} \nonumber \\ 
&& + S_{m}[g,\Psi],
\end{eqnarray}
where $\kappa =8\pi G$, $\Psi$
denotes any additional matter fields, $R_{\mu\nu}$ is the symmetric Ricci
tensor constructed with $\Gamma$. The term insight the root denote the determinant.
%which depends of an auxiliary metric $q_{\mu\nu}=g_{\mu\nu}+\kappa R_{\mu\nu}$. 
Here the matter is added in the usual way.
%and $R$ is the standard curvature scalar constructed with the usual 
%connection $\Gamma (g)$. 
%This action can reproduce the Eddington limit at
%large values of $\kappa R$ and Einstein limit at small values.
%, as we shall see in this paper.
%It is important to see that (\ref{palatini}) reproduces Einstein gravity
%within the vacuum. Two aspects that has been studied of this theory are in
%regions of high density: for a black hole metric and in the very early
%universe, given a proposal where the singularity can be avoided, pointing
%to an alternative theory of the big bang.
The connection between (\ref{palatini}) and cosmological observations has been done in \cite{Scargill:2012kg}. 
Despite its preliminaries success, in bouncing cosmological solutions cases it has already been observed that EiBI suffers from
instabilities associated with the growth of tensor perturbations \cite{EscamillaRivera:2012vz}. In latest works,
further considerations about the tensor perturbations in EiBI were made \cite{Avelino:2012ue,BeltranJimenez:2017uwv,BeltranJimenez:2017doy,Jana:2017ost}.
Moreover, the aim of this paper is to take a step forward in order to calculate the fluctuations of the EiBI tensor perturbations
and compute the graviton mass at two limits: for low-energy densities (General Relativity -GR-) and high-energy densities (Eddington limit).
%in two limits: Eddington and General Relativity (GR).
%: in \cite{Avelino:2012ue} were considered bouncing solutions
%to avoid this instability pathology and in \cite{BeltranJimenez:2017uwv} was presented the evolution of GW
%for non-singular solutions.

%In this work we revised the general formulation for the gravitational waves in 
%EiBI theory. Also we will see that the general asymptotic wave equations show the
%smooth transition between GR and Eddington regimes. 
This paper is organised as follows: In Sec. \ref{sec:EiBIintro} we will review
the field equations for the EiBI theory. In Sec. \ref{sec:gwintro} we summarise the limits in this theory. 
%and the limits of the EiBI theory in the non-perturbative limit are presented.  
In Sec. \ref{sec:EIBIGW} we calculate the GW equation for the EiBI theory and it will be thoroughly discussed. Also, as a main goal of this paper, we compute the evolution
of the graviton mass at both, high and low densities. In Sec. \ref{sec:EIBI-fluc} we explore the
fluctuation spectrum in this theory.

%-------------------------------------------------------------------------------------------------------------------------------------------------
%-------------------------------------------------------------------------------------------------------------------------------------------------

\section{EiBI field equations}\label{sec:EiBIintro}

%In this section we review some key features of \cite{Banados:2010ix} and
%introduce the procedure that gives to us the way to compute the tensor
%perturbations using the auxiliary metric $q_{\mu\nu}$ and the variation
%of the equations of motion that define the Eddington model.

From (\ref{palatini}) 
%the metric and the connection are
%treated as independent variables, this approach is known as Palatini 
%variation from where 
we can calculate the Einstein field equations by varying with respect to the metric $g_{\mu\nu}$ 
and the variation with respect to the connection fixes the affine connection to be $\Gamma$:
%The resulting field equations 
%of the EiBI theory 
%are, respectively
\begin{equation}\label{evolution1}
\sqrt{\left|\frac{q}{g}\right|}(q^{-1})^{\mu\nu} -\lambda g^{\mu\nu}=-\kappa T^{\mu\nu}, 
\end{equation}
%\sqrt{g}(q^{-1})^{\mu\nu} -\lambda \sqrt{g} g^{\mu\nu}=-\kappa \sqrt{g}T^{\mu\nu}, 
\begin{eqnarray}\label{evolution2}
&&\nabla_\alpha \left[\sqrt{q}(q^{-1})^{\mu\nu}\right]- \nabla_{\beta}\left\{\sqrt{q}
\left[\delta^{\mu}_{\alpha}(q^{-1})^{\beta\nu}
%\right.\right. \nonumber\\ && \left.\left.
+\delta^{\nu}_{\alpha}(q^{-1})^{\beta\mu}
\right]\right\}=0, \nonumber
\end{eqnarray}
\begin{equation}\label{evolution3}
q_{\mu\nu}=g_{\mu\nu}+\kappa R_{\mu\nu},
\end{equation}
where 
%$\tilde{q}=q^{-1}$, 
$\lambda=1+\kappa\Lambda$, and $\kappa$ is a constant with the inverse 
dimensions of $\Lambda$. Notice that these field equations are obtained from independent variation of the metric
and $\Gamma$.
%The main differences in comparison to the standard GR are that 
The auxiliary tensor $q_{\mu\nu}$ is not the space-time metric and $\lambda$ can be related to the cosmological
constant term from a GR point of view.

\section{Limits in the EiBI theory} \label{sec:gwintro}
Focusing on the dynamics of homogeneous and isotropic metric in (\ref{evolution1})-(\ref{evolution3}),
we consider a line element with time and spatial components 
for each metric:
\begin{eqnarray}\label{eq:metrics1}
 g^{00}&=&1, \hspace{0.8cm} g^{ij}=a^{-2}\delta^{ij}, \nonumber \\
  q^{00}&=&X^{-2}, \quad q^{ij}=(aY)^{-2}\delta^{ij}, 
\end{eqnarray}
%\begin{eqnarray}\label{eq:metrics2}
% q^{00}&=&X^{-2}, \quad q^{ij}=(aY)^{-2}\delta^{ij}, 
%\end{eqnarray}
%where $(aY)=aY$ is in conformal space-time. 

Eqs. (\ref{evolution1})-(\ref{evolution2}) can be solved analytically using 
%the metric
(\ref{eq:metrics1}) to derive the conventional Friedmann cosmology at late-times. 
%The trivial case is when we consider the non-perturbed equations, 
The zero-component evolution equation with $\sqrt{|q/g|}= |XY^{3}|$ is: 
\begin{equation}\label{eq:friedmann}
 3\kappa \left(H +\frac{\dot{Y}}{Y}\right)^2 =X^2 \left(1-\frac{3}{2Y^2}\right)+\frac{1}{2},
\end{equation}
where
\begin{eqnarray}\label{eq:X-Y}
 |X|&=&\frac{(1+\kappa P_{T})^{2}}{[(1+\kappa \rho_{T})(1-\kappa P_{T})^{1/4}]},\\
  |Y|&=&[(1+\kappa \rho_{T})(1-\kappa P_{T})^{1/4}],
\end{eqnarray}
%\begin{equation}\label{eq:Y}
% |Y|=[(1+\kappa \rho_{T})(1-\kappa P_{T})^{1/4}],
%\end{equation}
with $\rho_{T}=\rho +\Lambda$ and $P_{T}=P-\Lambda$. 
%For instance, the variables are in physical time. 
Let us assume radiation domination as: $\rho_{T}=\rho$
and $P_{T}=P=\rho/3$, we find that $X$ and $Y$ at late times behaves as:
\begin{eqnarray}
 |X|&\simeq& 1-\frac{5}{6}\kappa\rho+O(\kappa^2), \\
  |Y|&\simeq& a+\frac{a}{6}\kappa\rho+O(\kappa^2).
\end{eqnarray}
%\begin{equation}
% |Y|\simeq a+\frac{a}{6}\kappa\rho+O(\kappa^2).
%\end{equation}
If $X=Y=1$ the latter reduces to the low-energy densities limit (GR limit). Now, considering high energy densities
(Eddington limit) $\rho\rightarrow \rho_{B}$, where the subindex $B$ 
indicates the existence of a minimum value for the scale factor \cite{Banados:2010ix} then 
the approximation for the variables $X$ and $Y$ are
\begin{eqnarray}
 |X|&=&\frac{\left(1-\frac{\bar{\rho}}{3}\right)^2}{[(1+\bar{\rho})\left(1-\frac{\bar{\rho}}{3}\right)]^{1/4}}, \\
  |Y|&=&[(1+\bar{\rho})\left(1-\frac{\bar{\rho}}{3}\right)]^{1/4},
\end{eqnarray}
%\begin{equation}
% |Y|=[(1+\bar{\rho})\left(1-\frac{\bar{\rho}}{3}\right)]^{1/4},
%\end{equation}
where we introduce $\bar{\rho}=\kappa\rho$. We see a critical point at $\bar{\rho}=\rho_{B}=3$.
Rewriting (\ref{eq:friedmann}) we obtain
\begin{eqnarray}\label{eq:friedmann1}
 3H^2 &=&\frac{1}{\kappa}\left[\bar{\rho} -1 +\frac{1}{3\sqrt{3}}\sqrt{(\bar{\rho} +1)(3-\bar{\rho})^3}\right] \nonumber \\
&& \times \left[\frac{(1+\bar{\rho})(3-\bar{\rho})^2}{(3+\bar{\rho}^2)^2}\right],
\end{eqnarray}
where for $\bar{\rho} \ll 1$ we have $H^2 \simeq \rho/3$. Eq.(\ref{eq:friedmann1}) has critical points 
for $H(\rho_{B})=0$ in a maximum density $\rho_{B} =0,-1,3$. Each critical point appear when 
$Y^2 =3X^2 /(2X^2 +1)$. Notice that each critical density has an analytical solution 
that corresponds to an expansion of the scale factor depending of the sign of $\kappa$ (see Figure 1):
\begin{itemize}
 \item When $\bar{\rho}$ ($\kappa\approx 1$), $X=Y=1$, then we have a minimum scale factor at $\dot{a}=0$ and the universe its stationary and has a
minimum size $a=a_{B}\approx 10^{-32}(\kappa)^{1/4}a_{0}$, where $a_{0}$ is the scale factor today.
This replace the usual Big Bang singularity of Einstein's model by a cosmic bounce.
 \item When $\bar{\rho}=3 (\kappa >0)$, $X=Y=0$ and the
solution is exponential-like $(a/a_{B})-1 \propto e^{t-t_{B}}$, which corresponds to a loitering solution.
%Later, we shall see that this value of $\bar{\rho}$ corresponds to a linear grow of the graviton.
 \item When $\bar{\rho}=1$ ($\kappa <0$), 
 %the value for each variable are: 
 $X=(3\cdot 4^3)^{1/4}/9$ and
$Y=(4^3)^{1/4}$, with solution $(a-a_{B})\propto |t-t_{B}|^2$, which corresponds to a bouncing solution.
\end{itemize}
%All evolutions are show in the Figure 1.

Given that the solution for the radiation is $\rho= \rho_0/a^4=\rho_0 /(a_B +\delta a)^4$, we can expand the density
around the small variation of $a$ ($\delta a$),
\begin{eqnarray}
\rho &\propto& \frac{\rho}{a^4_B} + 4\rho_0 \frac{\delta a}{a^5_B} + O(\delta a^2), \\
\rho &\propto& \rho_B + 4\rho_B \frac{\delta a}{a_B} + O(\delta a^2),
\end{eqnarray}
where $\rho_B =\rho_0 a^{-4}_B$ is the maximum density. As $a=a_B +\delta a$ then $(a/a_B)-1 =\delta a/a_B$,
\begin{eqnarray}\label{eq:density-scale}
\rho &\propto& \rho_B + 4\rho_B \left(\frac{a}{a_B}-1\right) + O\left[a^{2}_B  \left(\frac{a}{a_B}-1\right)^2\right], \nonumber \\
a &\propto& 1+(t-t_B) + O[(t-t_B)^2]. 
\end{eqnarray}
%and for the scale factor:
%\begin{equation}\label{eq:scale}
%a \propto 1+(t-t_B) + O[(t-t_B)^2].
%\end{equation}
At early times (\ref{eq:density-scale}) shows a universe with a maximum density and constant scale factor.

\begin{figure}[t]
\begin{center}
\includegraphics[width=8cm]{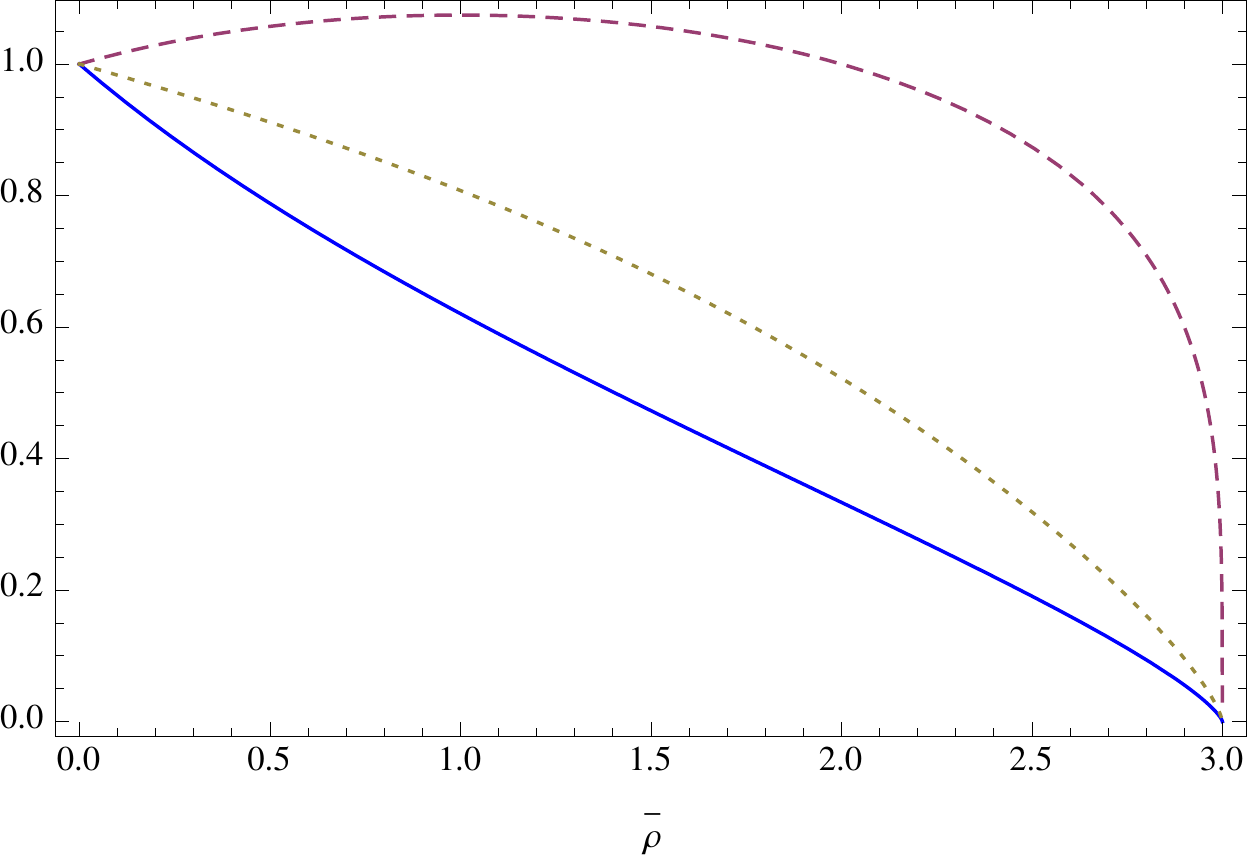}
\caption{The evolution of $X$ (blue solid line) and $Y$ (purple dashed line) from the
low-energy to high-energy density limit. We observe two critical points for each limit at 
$\rho\rightarrow\rho_{B}=0$ and $\rho\rightarrow\rho_{B}=3/\kappa$, respectively. The 
yellow-dotted curve represents the evolution of $Y^2 =\frac{3X^2}{2X^2 +1}$ that gives
$H^2 =0$.}
\end{center}
\end{figure}

%-------------------------------------------------------------------------------------------------------------------------------------------------
%-------------------------------------------------------------------------------------------------------------------------------------------------

\section{EiBI tensor perturbations} \label{sec:EIBIGW} 
We can consider a perturbed homogeneous and isotropic spacetime by choosing the two metrics to be
of the form:
%Now, we consider a line element with a perturbation on the metrics $g$ and $q$ as: 
\begin{eqnarray}\label{eq:metrics2}
 g^{00}&=&1, \hspace{0.8cm} g^{ij}=a^{-2}(\delta^{ij}-h^{ij}), \nonumber \\
  q^{00}&=&X^{-2}, \quad q^{ij}=(aY)^{-2}(\delta^{ij}-\gamma^{ij}), 
\end{eqnarray}
%\begin{eqnarray}\label{eq:metrics2}
% q^{00}&=&X^{-2}, \quad q^{ij}=(aY)^{-2}(\delta^{ij}-\gamma^{ij}), 
%\end{eqnarray}
where $h_{ij}$ and $\gamma_{ij}$ are traceless and transverse, i.e $\partial_i h^{ij}=\partial_i\gamma^{ij}=0$,
$h_{ii}=\gamma_{ii}=0$, respectively.
To construct the perturbed field equations we compute the quantities:
\begin{eqnarray}
(q^{-1})^{ij} &=&(aY)^{-2} (\delta^{ij}-\gamma^{ij}), \\
(g)^{ij}&=&a^2 (\delta^{ij} -h^{ij}),  \\ 
\delta T^{ij} &=&-Pa^{-2}h^{ij},
\end{eqnarray}
where we take $T_{\mu\nu}=(\rho+P)u_{\mu}u_{\nu} +Pg_{\mu\nu}$ with $u^{\mu}=(1,0,0,0)$. 

An interesting results derived from the field equations is that $\gamma_{ij}=h_{ij}$, i.e it was found in \cite{EscamillaRivera:2012vz}
that $\gamma_{ij}$ is completely locked to the behaviour of $h_{ij}$. After following this consideration
we can write the evolution equation for $h_{ij}$ as
%The general equation for $h$ can be compute by taking the derivative of the relationship between $\gamma$ and $h$. This expression for
%$\delta R_{ij}$ plus the second term of the l.h.s of (\ref{eq:ricci2}) gives
\begin{eqnarray} \label{eq:grav}
&& h''_{ij} + \left(4\frac{(aY)'}{(aY)} -2\frac{a'}{a}\right) h'_{ij}+\left[\frac{X''}{X} +2\frac{X'}{X}\frac{(aY)'}{(aY)} +\frac{(aY)''}{(aY)}
\right. \nonumber\\ &&
\left. -4\frac{X'}{X^2}\frac{(aY)'}{(aY)}\frac{a'}{a} +2\frac{a'^2}{a^2} -\frac{a''}{a} -2\frac{(aY)'^2}{(aY)^2} -2\frac{X'^2}{X^2} 
\right. \nonumber\\ &&
\left.-2k^2 \frac{X'}{X}\frac{(aY)'}{(aY)}  +k^2 +\frac{X(aY)^3 +a^3\lambda}{\kappa\lambda a X^2 (aY)^2}\right] h_{ij} =0.
\end{eqnarray}
where the prime denotes derivatives w.r.t the conformal time $\eta$. This \textit{graviton} equation can be rewriten by using $|aY|$ and the component $R_{00}$ to obtain
\begin{eqnarray}\label{eq:wave-rho}
&& \left[\kappa\sqrt{-(\kappa\rho+1)(\kappa\rho-3)^3}(\kappa\rho +1)(\kappa^2 \rho^2 +3)\right]h''_{ij} \nonumber \\ &&
+\left[-\frac{2}{9}\sqrt{3}\kappa a(\kappa\rho -1)(\kappa\rho -3)\sqrt{-(\kappa\rho)(\kappa\rho -3)^3} \right. \nonumber \\ && \left.
\sqrt{(9\kappa\rho +\sqrt{3}\sqrt{(\kappa\rho+1)(\kappa\rho -3)^3}-9)(\kappa\rho +1)}\right]h'_{ij} \nonumber \\ &&
+\{ (\kappa^2 \rho^2 +3)\sqrt{-(\kappa\rho +1)(\kappa\rho -3)^3} \left[\frac{2}{3}a^2  \right. \nonumber \\ && \left.
\sqrt{-(\kappa\rho +1)(\kappa\rho -3)^3}+k^2 \kappa(\kappa\rho -3)\right] \nonumber \\ &&
+\frac{2\sqrt{3}}{3}(\kappa\rho +1)(\kappa\rho -3)^3 \}h_{ij}=0.
\end{eqnarray}

At low-energy densities, if we expand the r.h.s of (\ref{eq:wave-rho}) we obtain
\begin{eqnarray}
&&\frac{1}{\sqrt{3}} \left(2\sqrt{\rho}h'_{ij} +\sqrt{3}k^2 h_{ij}\right) +4\left(\sqrt{3}\rho^{3/2} h'_{ij} +\frac{k^2}{3}\rho h_{ij}\right)\kappa 
\nonumber \\ &&+O(\kappa^2) \approx 0,
\end{eqnarray}
where at late times $(\kappa\ll 1)$ and using $3H^2 =\rho$ we recover the Helmholtz equation.

At high-energy densities, (\ref{eq:wave-rho}) has a critical point $\rho_B =3/\kappa$, therefore 
\begin{equation}\label{eq:edd}
h''_{ij} =0 \quad \rightarrow \quad h_{ij}\approx h_0 \eta,
\end{equation}
then $h_{ij}$ grows linearly at early times for 
\begin{equation}
\displaystyle{\lim_{\bar{\rho} \to 3}} \left[\kappa \sqrt{-(\kappa\rho+1)(\kappa\rho-3)^3} (\kappa\rho+1)(\kappa^2\rho^2 +3)\right]=0.
\end{equation}

%\begin{figure}[t]
%\begin{center}
%\includegraphics[width=7cm]{rho.png}
%\caption{Evolution of $h''_{ij}(\bar{\rho})$. We notice that $h''$ approaches 
%to zero at high-energy density limit $\bar{\rho}\rightarrow 3/\kappa$ for $|\kappa|=1$.}
%\end{center}
%\label{t1}
%\end{figure}

\begin{figure}[t]
\centering
\includegraphics[width=0.46\textwidth,origin=c,angle=0]{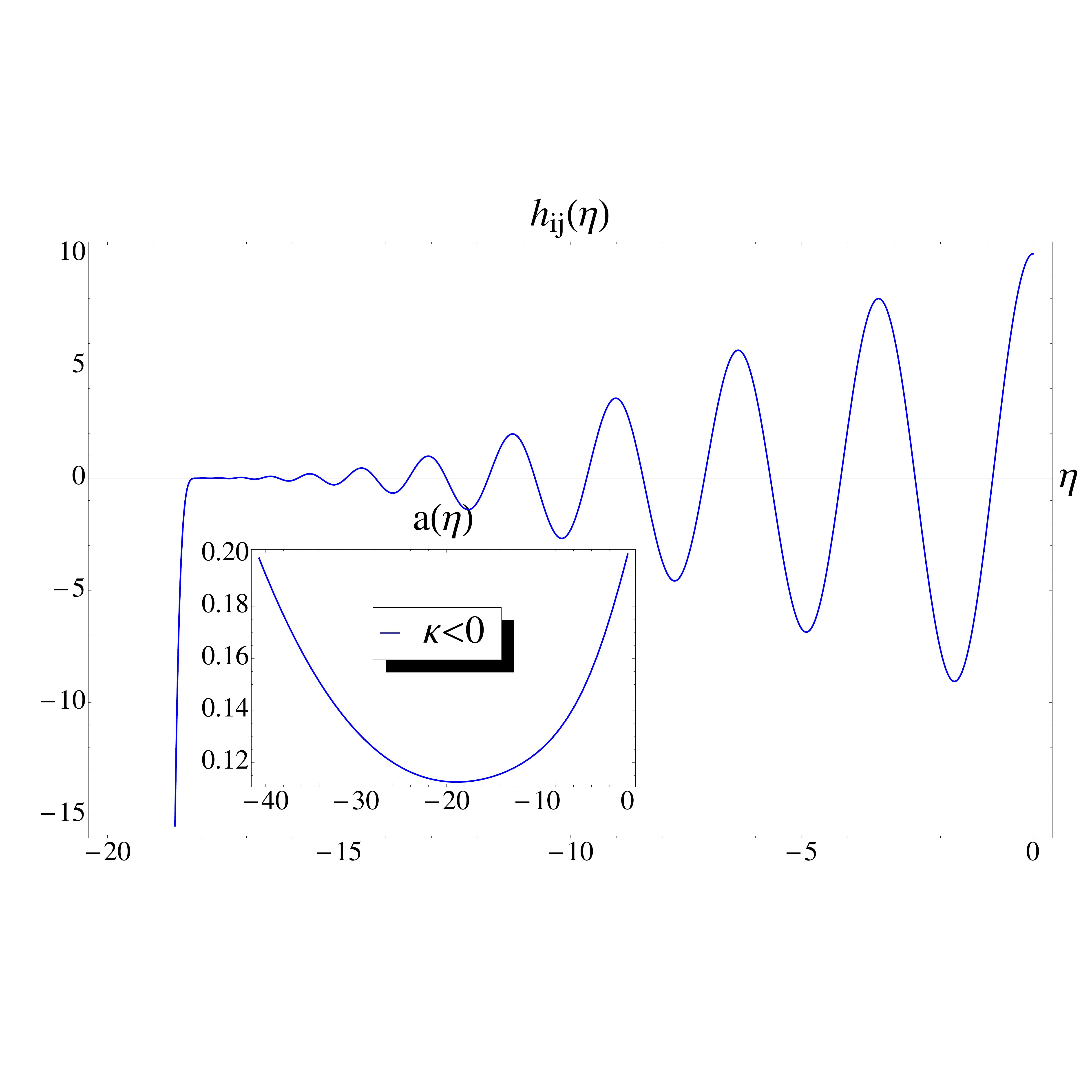} 
\includegraphics[width=0.46\textwidth,origin=c,angle=0]{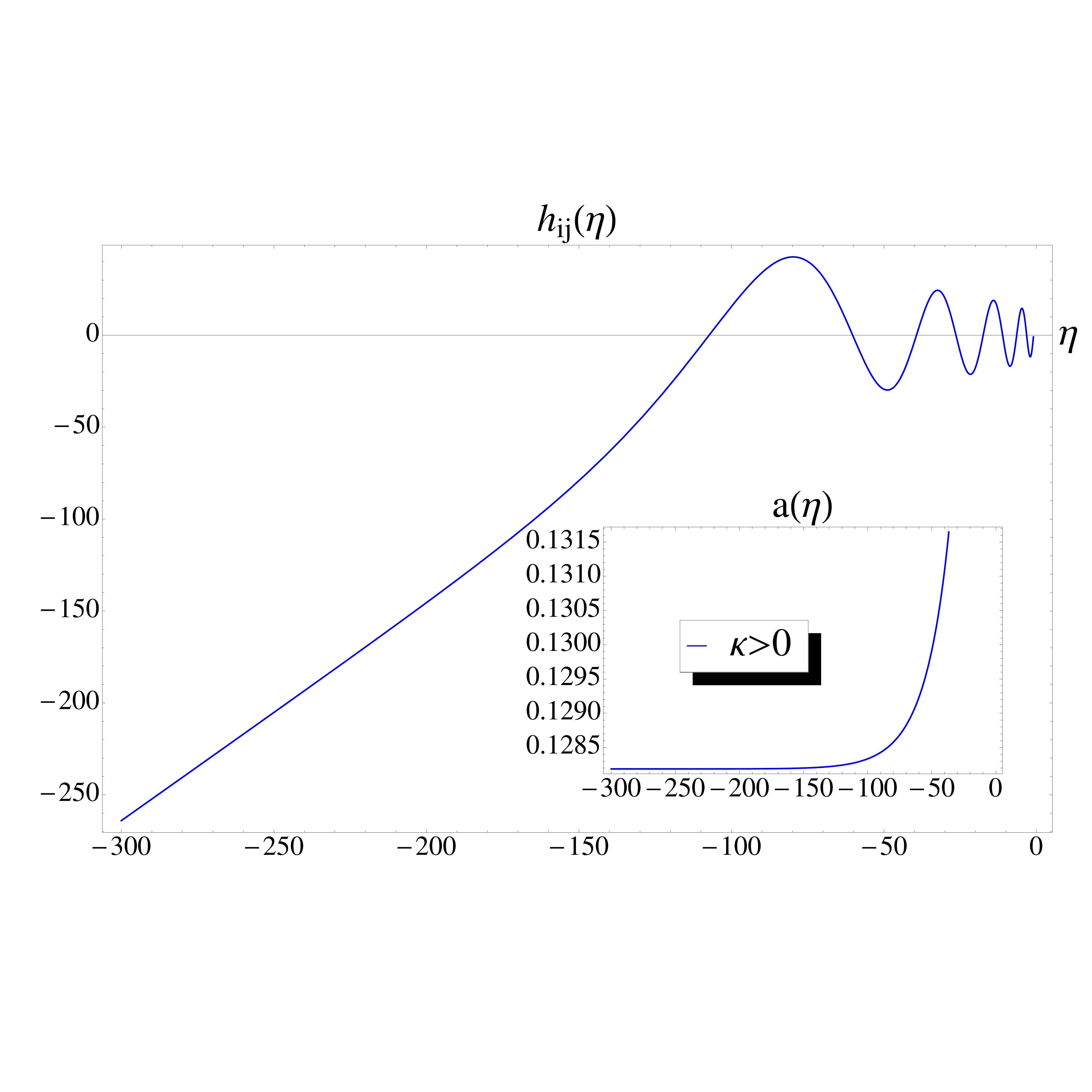}
\caption{Numerical solutions for (\ref{eq:grav}) in 
conformal time. The footnote plots (inside) show the evolution of the scale factor for both values of $\kappa$.}
\end{figure}

\begin{figure}[t]
\begin{center}
\includegraphics[width=7.4cm]{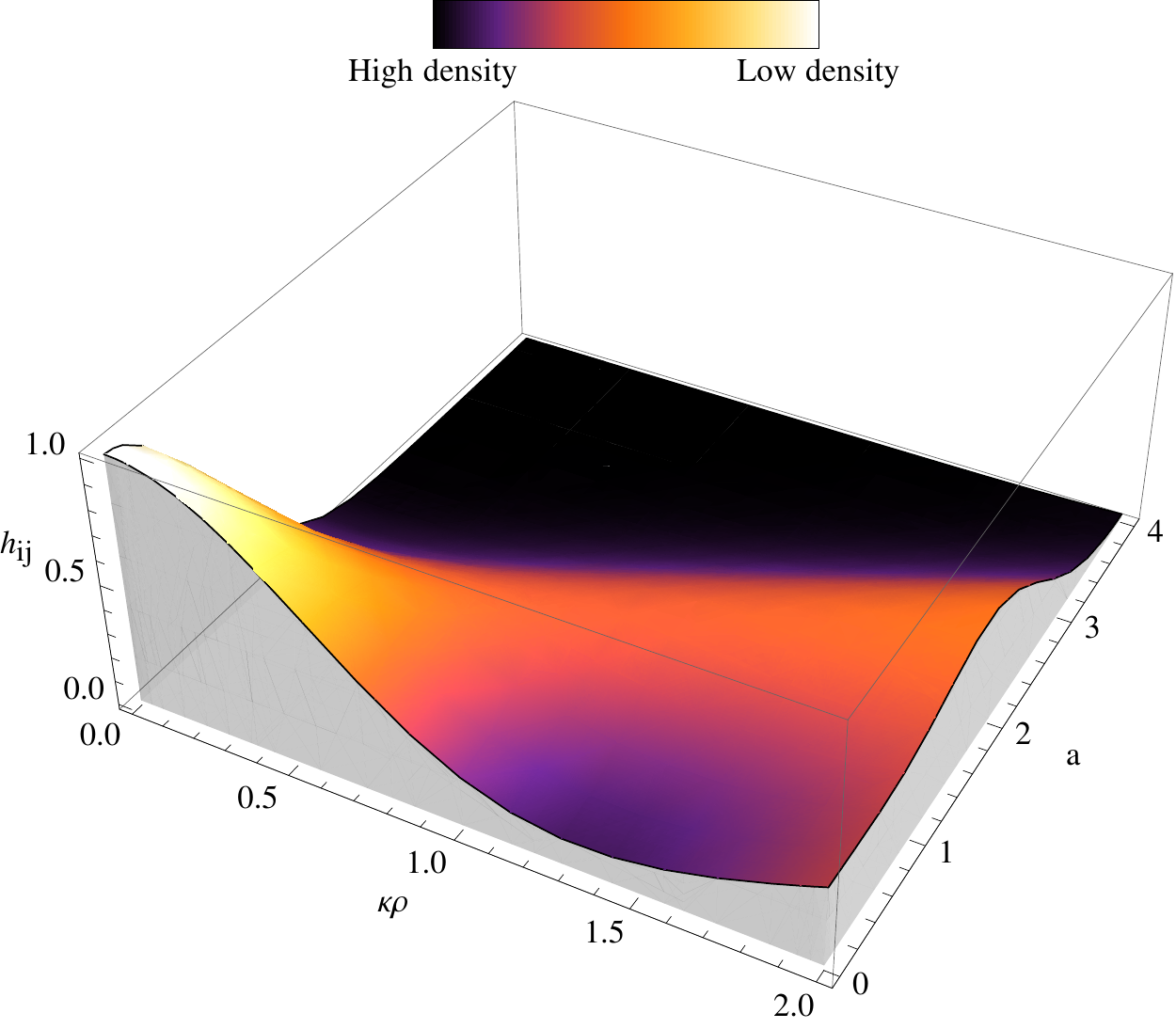}
\caption{Numerical solution for (\ref{eq:wave-rho}). We notice the transition between low-energy and high energy densities.}
\end{center}
\end{figure}

Performing the numerical integration of (\ref{eq:grav})-(\ref{eq:wave-rho}) we see the predicted behaviour (Figure 2): From the evolution of the scale factor we notice the 
smooth transition between high-energy densities (where the EBI dynamics plays its role) and low-energy densities (GR) (Figure 3). From the solution $h_{ij}$, we notice the linear grow of the GW  (\ref{eq:edd}) and
as time evolves we have the damped oscillations in the GR limit.

\subsection{Graviton mass}
Consider the field equations 
\begin{equation}
R^{\mu\nu} -\frac{1}{2}g^{\mu\nu}R =-\kappa (T^{\mu\nu} +T^{\mu\nu}_{\text{mass}}),
\end{equation}
where usually the extra term $T^{\mu\nu}_{\text{mass}}$ depends of the graviton mass $m_g$ and 
the background metric
\begin{eqnarray}
T^{\mu\nu}_{\text{mass}} &=& -m_g \left\{ (g^{-1}_0)^{\mu\sigma} \left[ (g-g_0)_{\sigma\rho} -\frac{1}{2}(g_0)_{\sigma\rho} (g_0)^{\alpha\beta} 
\right. \right. \nonumber \\ && \left.\left.
(g-g_0)_{\alpha\beta}\right] ((g_0)^{-1})^{\rho\nu} 
\right\},
\end{eqnarray}
where if $m_{g} \rightarrow 0$ we recover the usual Einstein field equations. If we consider the background metric with a small perturbation 
to obtain for this mass term:
\begin{eqnarray}
T^{\mu\nu}_{\text{mass}}  = -m_g \left\{ h_{\mu\nu} -\frac{1}{2} \left[((g_0)^{-1})^{\alpha\beta} h_{\alpha\beta}\right](g_0)_{\mu\nu}\right\},
\end{eqnarray}
where, for $\delta T_{\mu\nu}=0$, $\delta G_{\mu\nu} = -\kappa\delta T^{\text{mass}}_{\mu\nu}$, we can rewrite the perturbed equation as:
\begin{equation}\label{eq:perturbation1}
h''_{ij} +2Hh'_{ij} +(k^2 +m^{2}_{g}) h_{ij}=0.
\end{equation}
This equation is similar to the equation for a free massive scalar field in a flat FRW background. Now, from the perturbation of (\ref{evolution3})
$\kappa\delta R_{ij} =\delta q_{ij}-\delta g_{ij}$ we obtain:
\begin{equation}
\delta R_{ij} +\frac{a^2}{\kappa}\left(\frac{XY^3}{\lambda}+1\right)h_{ij}=0.
\end{equation}
%According to this, 
If we compare the latter with (\ref{eq:perturbation1})
we obtain
\begin{equation}
m^{2}_{g} =\frac{1}{\kappa} \left[\frac{(1+\kappa\rho_T)(1-\kappa P_T )^3}{1+\kappa\Lambda}+1\right],
\end{equation}
which is the graviton mass that takes the following values:
\begin{itemize}
\item $\frac{XY^3}{\lambda} < -1$ gives a large tachyonic mass $m^{2}_{g}<0$ implying the unstable evolution of tensor perturbations,
\item $\frac{XY^3}{\lambda} > -1$, the growth of the tensor perturbations is suppressed.
\end{itemize}
Eq. (\ref{eq:perturbation1}) reduces to
\begin{equation}
h''_{ij} +2Hh'_{ij} +k^2 h_{ij} =0, \quad  \text{for} \quad p^2 \gg m^{2}_{g},
\end{equation}
and 
\begin{equation}
h''_{ij} +2Hh'_{ij} +m^{2}_{g} =0, \quad p^2 \ll m^{2}_{g}, \rightarrow \lambda\left(\frac{\kappa k^2}{a^2}-1\right)\ll XY^3,
\end{equation}
where $p^2 \equiv a^{-2}k^2$ is the physical momentum.

\begin{figure}[t]
\begin{center}
\includegraphics[width=6.cm]{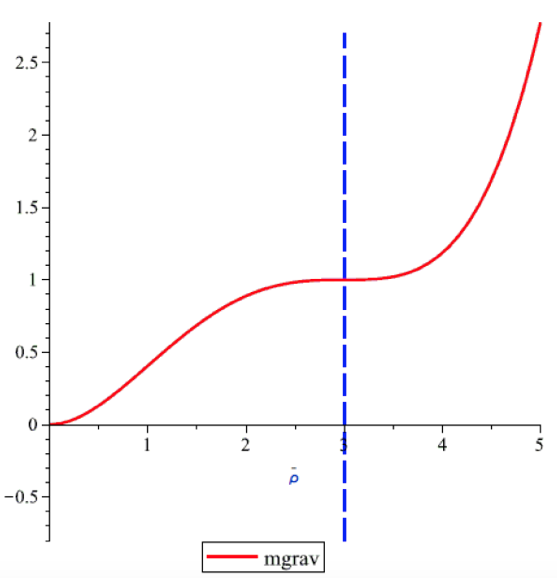}
\caption{Evolution of the graviton mass in terms of $\bar{\rho}$. The dashed-blue
line divide the high-energy density limit (right side) and low-energy density limit (left side). The critical point is
located at $\bar{\rho}=3/\kappa$ for $|\kappa|=1$. 
%We normalize the scale factor $a$ by the minimum length $a_{B}$.
}
\end{center}
\label{t1}
\end{figure}

At high-energy densities ($\kappa\rho \rightarrow 3$) and in radiation regime (Figure 4)
\begin{equation}
m^{2}_{g_{\text{rad}}} \propto \frac{1}{\kappa} \left[(1+\bar{\rho})\left(1-\frac{1}{3}\bar{\rho}\right)+1\right] =\frac{1}{\kappa}.
\end{equation}
%The evolution of this graviton is show in Figure 5. 
When $m_g >1$ there is a growth
of the tensor perturbations, after the graviton crosses the critical point $\bar{\rho} = 3/\kappa$ at low-energy density ($0< m_g <1$) the growth
is suppressed.
% as the density decreases. 

%-------------------------------------------------------------------------------------------------------------------------------------------------
%-------------------------------------------------------------------------------------------------------------------------------------------------

\section{Fluctuations in GW EiBI} \label{sec:EIBI-fluc}

%The theory of cosmological perturbations relies essentially on the assumption 
%that the background is described by pure classical GR, while the perturbations themselves 
%are derived from quantum perturbations. 
%It is at semiclassical approach, where the background 
%is classical and the perturbations are quantized, and the fact that the background satisfies 
%Einstein's equations is usually used in the simplification of these equations. 
%Therefore, 
We 
rewrite the graviton equation (\ref{eq:grav}) as:
\begin{equation}\label{eq:wavepert}
 h''_{ij}+Fh'_{ij}+(G+Jk^2)h_{ij}=0, 
\end{equation}
where
\begin{equation}
 F=2\left[2\frac{(aY)'}{(aY)}-\frac{a'}{a}\right],
\end{equation}
\begin{eqnarray}
 G&=&\frac{X''}{X}+2\left(\frac{X'}{X}\right)\left[\frac{(aY)'}{(aY)}\right] +\frac{(aY)''}{(aY)} \nonumber \\&&
-4\left(\frac{X'}{X^2}\right)\left[\frac{(aY)'}{(aY)}\right]\left(\frac{a'}{a}\right)
+2\left(\frac{a'}{a}\right)^2 
 \nonumber \\ &&
 -\frac{a''}{a} -2\left(\frac{X'}{X}\right)^2 -2\left[\frac{(aY)'}{(aY)}\right]^2
+\xi ,
\end{eqnarray}
\begin{equation}
 J=1-2\left(\frac{X'}{X}\right)\left[\frac{(aY)'}{(aY)}\right],
\end{equation}
\begin{equation}
 \xi =\frac{X(aY)^3 +a^3 \lambda}{\kappa\lambda a X^2 (aY)^2}.
\end{equation}

%In \cite{Lidsey:1995np}
%which assume the validity of the Einstein-Hilbert action (at quadratic order in $h_{\mu\nu}$), 
%it was shown that such simple equations for quantum linear cosmological perturbations can 
%also be obtained without ever using any equations for the background, instead the authors employed
%a Fourier decomposition of the gravitational waves. This can be accomplished through a canonical 
%change of variable and after consider this as a quantum operator is possible to obtain a field
%equation with the same characteristic as the damped oscillator equation. These results open the 
%way to also quantise the background, and use these simple equations to evaluate the evolution
%of the quantum linear perturbations on it.
We use the change of variable $u=zh$ to avoid the \textit{friction-type} term. We have at low-energy densities:
\begin{equation}\label{eq:friedmannz}
 u''+\left(-\frac{z''}{z}+k^2 \right)u=0,
\end{equation}
%for GR regime 
and at high-energy densities:
\begin{equation}\label{eq:eddingtonz}
 u''+\left(-\frac{z''}{z}+G+Jk^2 \right)u=0,
\end{equation}
%for ER.

For each Fourier mode $k$ the above equations are the harmonic oscillator with time-dependent 
frequency
\begin{eqnarray}\label{eq:def1}
 \omega_{k}&:=&\sqrt{k^2 -\frac{z''}{z}}, \\
  \omega_{k}&:=&\sqrt{G +Jk^2 -\frac{z''}{z}}.
\end{eqnarray}
%\begin{equation}
% \omega_{k}:=\sqrt{G +Jk^2 -\frac{z''}{z}}.
%\end{equation}
If the expansion is rapid enough, $\omega_k$ becomes imaginary.

Will use the following definition for the quantum fluctuations where $g_{B}$ is an
observable with mean value 
\begin{equation}
 g_B =<\Omega|g_B |\Omega> =<0| Aa +Ba^{\dag} |0>=0,
\end{equation}
and the variance is 
\begin{equation}
 [\delta g_B]^2 =<\Omega|{g_B}^2 |\Omega> \approx a^{-2}k^3 |u_k|^2.
\end{equation}
We define the fluctuation spectrum as the standard deviation as a function of $k$
\begin{equation}\label{eq:def2}
 \delta g_k :=a^{-1} k^{3/2}|u_k|.
\end{equation}

%--------------------------------------------------------------------------------
%--------------------------------------------------------------------------------

\subsection{Solutions of Eq.(\ref{eq:friedmannz})}

\begin{itemize}
 \item Case 1. Minkowski spacetime $z=1$. Solving (\ref{eq:friedmannz}) and using (\ref{eq:def1})-(\ref{eq:def2}), the solutions for the mode function and the fluctuation spectrum are
\begin{eqnarray}\label{eq:case1-1}
 u_{k,\eta}&=&(1/\sqrt{k}) e^{i\eta k}, \nonumber \\
  \delta g_k &=&k^{3/2}/\sqrt{k} =k^3.
\end{eqnarray}
%\begin{equation}\label{eq:fluc1}
% \delta g_k =k^{3/2}/\sqrt{k} =k^3.
%\end{equation}
We observed that at small $k$ (large scale), the fluctuations are strongly suppressed. 
%The evolution is show in Figure 6, where
%we consider different values of $k$ as $k_{1}\ll k_{2}\ll k_{3}$. The blue-line represents a late-time universe, 
Analogous to the
numerical solutions, when the scale factor is constant we observed a fast growing of the graviton mode $h_{ij}$.
%\begin{figure*}
%\centering
%\includegraphics[width=16.0cm]{friedmannspectrum.pdf}
%\caption{Evolution of (\ref{eq:fluc1}) for the Case 1: Minkowski spacetime with $z=1$ in (\ref{eq:friedmannz}). The blue-line represent a late-time universe.}
%\label{fig:y_best_minimum2}
%\end{figure*}

\item Case 2. De Sitter spacetime. We consider
$ z=e^{\alpha t}=-(\alpha\eta)^{-1},$
with $\eta (t)= \int{z(t')^{-1}dt'}$. The frequency is 
\begin{equation}
 \omega_{k,\eta}=k^2 -\frac{2}{\eta^2}.
\end{equation}
The modes $k$ oscillate if $|\eta| \gg \sqrt{2}/k$ and the $\omega$ is an imaginary 
quantity when $|\eta| \ll \sqrt{2}/k$. The solutions are
\begin{eqnarray}
  u_{k,\eta}&=&-\sqrt{\frac{2}{\pi k^3}} \left[(C_1 k\eta +C_2)\cos{(k\eta)}  \right. \nonumber\\&&\left.
  -(-C_1 +C_2 k\eta)\sin{(k\eta)}\right],
\end{eqnarray}
and
\begin{eqnarray}\label{eq:fluctuation}
 \delta g_{k,\eta}&=&-\alpha\eta\sqrt{\frac{2}{\pi}}\left[(C_1 k\eta +C_2)\cos{(k\eta)} \right. \nonumber\\&&\left.
 -(-C_1 +C_2 k\eta)\sin{(k\eta)}\right], 
\end{eqnarray}
with $C_1$ and $C_2$ constants of integration. 
Notice that as $t\rightarrow -\infty$ we have $\eta\rightarrow -\infty$, but as $t\rightarrow \infty$ 
we have $\eta\rightarrow 0$\footnote{This last condition is only correct if the integral constant of $\eta$ vanishes.}. 
At large $k$ we have the usual fluctuation spectrum for Minkowski spacetime. As $t\rightarrow \infty$, the
fluctuations vanishes during the De Sitter expansion.

%\begin{figure*}[htbp]
%\centering
%\includegraphics[width=19.0cm]{desitter_case.pdf}
%\caption{}
%\label{fig:y_best_minimum2}
%\end{figure*}

\end{itemize}

%--------------------------------------------------------------------------------
%--------------------------------------------------------------------------------

\subsection{Solutions for Eq.(\ref{eq:eddingtonz})}

For (\ref{eq:eddingtonz}) is not so simple to consider the same assumptions as the latter case since there
is a dependence of $X$ and $aY$ in $G$ and $J$. Therefore, let us consider an expansion over $1/\eta$.
We can rewrite the expressions for $X$ and $Y$ as:
\begin{eqnarray}\label{eq:Y}
 |Y|&=&[(1+\kappa \rho_{T})(1-\kappa P_{T})^{1/4}], \\
  |X|&=&\frac{(1+\kappa P_{T})^{2}}{|Y|},
\end{eqnarray}
%\begin{equation}\label{eq:X}
% |X|=\frac{(1+\kappa P_{T})^{2}}{|Y|},
%\end{equation}
where with $\rho =r_0 -\frac{\lambda}{\kappa}$. The expressions for the total density
and pressure are
\begin{eqnarray}
 P_T &=&\frac{1}{\kappa}(1-\lambda\pi_0)-\pi_0 \rho =-\frac{1}{\kappa}\pi_0 r_0, \\
  \rho_{T}&=&r_0 -\frac{1}{\kappa},
\end{eqnarray}
%\begin{equation}
% \rho_{T}=r_0 -\frac{1}{\kappa},
%\end{equation}
and now
\begin{eqnarray}
 |X|&=&\frac{1-\pi_0 r_0}{\kappa r_0 (1+\pi_0 r_0)^{1/4}}, \\
  |Y|&=&\kappa r_0 (1+\pi_0 r_0)^{1/4}.
\end{eqnarray}
%\begin{equation}
% |Y|=\kappa r_0 (1+\pi_0 r_0)^{1/4}.
%\end{equation}
In the asymptotic past these expansions are reduced to $p_{1}=\pi_0$ and $\rho_1 =r_0$. 
We rewrite (\ref{eq:eddingtonz}) for the ER as:
\begin{equation}
 u'' +\left(-\frac{z''}{z} +\xi_0 +k^2 \right)u=0,
\end{equation}
where
\begin{equation}
 \xi_0=\frac{(1-\pi_0 r_0)(1+\pi_0 r_0)^{1/2}-\lambda\kappa r_0}{\kappa r_0 \left[\kappa\lambda (1-\pi_0 r_0)(1+\pi_0 r_0)^{1/4}\right]}.
\end{equation}
%Now is possible to analize the cases above discussed, but the behaviour it will depend of the values of $\pi_0$ and $r_0$.
For the Minkowski case, we obtain similar solutions as in (\ref{eq:case1-1}) but with an extra constant term $\xi_0$ in the exponential.
For the DeSitter case also we obtain oscillating solution, but the harmonic functions will be weighted by a $(\sqrt{\xi_0} + k)$ term. When $\pi_0, r_0 \ll 1$ there is an increase on the amplitude of the fluctuation spectrum. In Figure 5 we show the power spectrum of the graviton equation ($P_{g}\propto |\delta g_k|^2$).

\begin{figure}[t]
\begin{center}
\includegraphics[width=8cm]{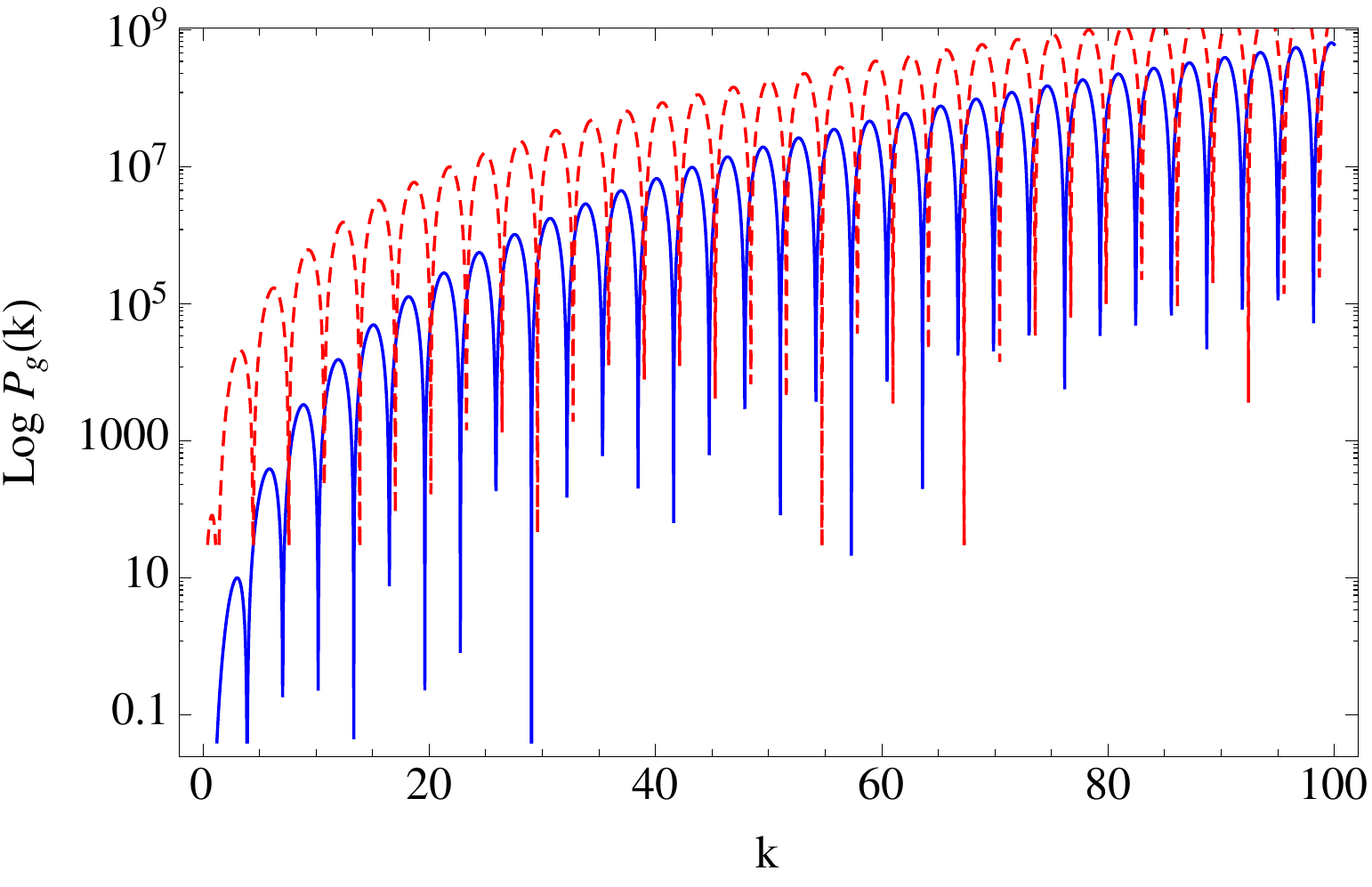}
\caption{Power spectrum for the GW (\ref{eq:eddingtonz}) using the fluctuations (\ref{eq:fluctuation}). The blue-solid curve represent a value of $\xi_0 =99$ and the
red-dashed curve for $\xi_0 =0.003$.} 
\end{center}
\label{t7}
\end{figure}
%-------------------------------------------------------------------------------------------------------------------------------------------------
%-------------------------------------------------------------------------------------------------------------------------------------------------

\section{Conclusions}
The EiBI theory has been a successful proposal for modify gravity theories, in which it is replaced the usual Big Bang 
singularity of Einstein's model by a cosmic bounce. Also, it was observed that this proposal suffers from
a tensor instability.
%The cosmology behind this theory has been done extensively from the classical point of view until
%perturbations, where, as we mentioned, there is a pathology in the tensor perturbations. 
In regards to this, here we have discussed the evolution of the EiBI-GW equation. Furthermore, we obtain the value
for the graviton mass in EiBI gravity at high and low energy densities, where for $k\ll1$ the
fluctuations are strongly suppressed and for $k\gg 1$ these ones vanish during the De Sitter expansion.

Work
still needs to be done before compare with current observations. 
Although within this paper, we review the importance of use the EiBI theory in future test of GW. 
%This proposal should be a phenomenological model for the dark sector of the universe.

%-------------------------------------------------------------------------------------------------------------------------------------------------
%-------------------------------------------------------------------------------------------------------------------------------------------------

\begin{acknowledgments}
I thank P.G. Ferreira and M. Ba\~nados for past discussions of these ideas. This work was financial supported by MCTP-UNACH.
\end{acknowledgments}

%-------------------------------------------------------------------------------------------------------------------------------------------------
%-------------------------------------------------------------------------------------------------------------------------------------------------

%-------------------------------------------------------------------------
%-------------------------------------------------------------------------

\end{document}